\documentclass[twocolumn,showpacs,preprintnumbers,amsmath,amssymb,prl]{revtex4}
\usepackage{graphicx}

\begin{document}

% ------------------------------------------------------------------------
\title{Transport spectroscopy of a single dopant in a gated silicon nanowire}
% ------------------------------------------------------------------------

\author{H. Sellier}
\altaffiliation[Present address: ]{Laboratoire de Spectrom\'etrie
Physique, Universit\'e Joseph Fourier, Grenoble, France}
\author{G. P. Lansbergen}
\author{J. Caro}
\author{S. Rogge}
\affiliation{Kavli Institute of Nanoscience, Delft University of
Technology, Lorentzweg 1, 2628 CJ Delft, The Netherlands}

\author{N. Collaert}
\author{I. Ferain}
\author{M. Jurczak}
\author{S. Biesemans}
\affiliation{InterUniversity Microelectronics Center (IMEC),
Kapeldreef 75, 3001 Leuven, Belgium}

% ------------------------------------------------------------------------

\begin{abstract}

We report on spectroscopy of a single dopant atom in silicon by
resonant tunneling between source and drain of a gated nanowire
etched from silicon on insulator. The electronic states of this
dopant isolated in the channel appear as resonances in the low
temperature conductance at energies below the conduction band
edge. We observe the two possible charge states successively
occupied by spin-up and spin-down electrons under magnetic field.
The first resonance is consistent with the binding energy of the
neutral $D^0$ state of an arsenic donor. The second resonance
shows a reduced charging energy due to the electrostatic coupling
of the charged $D^-$ state with electrodes. Excited states and
Zeeman splitting under magnetic field present large energies
potentially useful to build atomic scale devices.

\end{abstract}

% ------------------------------------------------------------------------

\date{\today}


\maketitle

% ------------------------------------------------------------------------

Dopant atoms are essential in semiconductor technology since they
provide extrinsic charges necessary to create devices such as
diodes and transistors. Nowadays the size of these electronic
devices can be made so small than the discreteness of doping can
influence their electrical properties \cite{asenov-99-nt}. On the
other hand it may be an important breakthrough if a dopant could
be used as the functional part of a device instead of just
providing charges. As an example, dopant-based spin qubits in
silicon are possible candidates for quantum computation
\cite{kane-98-nat,vrijen-00-pra} thanks to their longer spin
coherence time \cite{tyryshkin-03-prb} as compared to
two-dimensional quantum dots defined by top gates in III/V
heterostructures \cite{elzerman-05-review,petta-05-sci}. Although
dopants are well known in bulk semiconductors, specific questions
arise in the context of nano-scale devices like the reduced
life-time of the two-electron state under electric field involved
by read-out schemes of spin qubits \cite{testolin-05-prb}. The aim
of this work is to study the electronic states of single dopants
in gated silicon nanostructures to bring information useful for
these issues.

Electron tunneling through isolated impurities has been observed
previously in two-terminal devices such as GaAs/AlGaAs double
barrier heterostructures \cite{geim-94-prl,deshpande-96-prl}. Here
we present experimental results on electron transport through the
localized states of individual n-type dopants in silicon
nanowires. In contrast to previous studies, our devices have a
three-terminal configuration with source, drain, and gate
electrodes allowing a detailed investigation of charge, orbital,
and spin states. In particular, we observe both the neutral $D^0$
and negatively charged $D^-$ states, and compare their binding
energy with the case of bulk dopants. This work provides a
quantitative description of the electronic properties of a single
dopant connected to electrodes in a gated nanostructure. This is
the first transport experiment measuring the charge states of a
real atomic system with a $1/r$ attracting Coulomb potential, thus
very different from quantum dots with harmonic potentials.

% ------------------------------------------------------------------------
\begin{figure}[b]
\includegraphics[width=8cm,clip,trim=0 0 0 0]{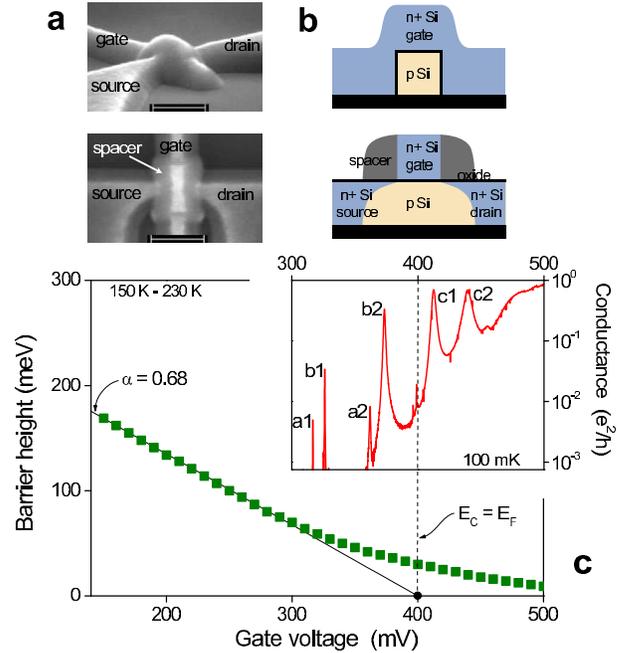}
\caption{(color online). (a) SEM images of typical devices, the
scale bar is 200\,nm. (b) Cross-sections showing n- and p-type
regions. The channel forms at the gate oxide with enhancement at
the top corners. (c) Height of the channel barrier measured at
high temperatures (squares). The band edge E$_\text{C}$ reaches
the Fermi level E$_\text{F}$ at 400\,mV. Inset: Low temperature
differential conductance showing resonances below this threshold
(a1 to b2) and above (c1 and c2).} \label{fig1}
\end{figure}
% ------------------------------------------------------------------------

Our devices are 60\,nm tall crystalline silicon wires (fins) with
large contacts patterned by 193\,nm optical lithography and dry
etching from Silicon-On-Insulator. After a boron channel
implantation, a 100\,nm poly-crystalline silicon was deposited on
top of a nitrided oxide (1.4\,nm equivalent oxide thickness), then
received a phosphorus (P) implant as pre-doping, and was patterned
using an oxide hard-mask to form a narrow gate
(Fig.\,\ref{fig1}\,a,b). Next, we used high-angle arsenic (As)
implantations as source/drain extensions, while the channel was
protected by the gate and 50\,nm wide nitride spacers and remains
p-type. Finally, As and P implants and a NiSi metallic silicide
are used to complete the source/drain electrodes. Conventional
operation of this n-p-n field effect transistor is to apply a
positive gate voltage to create an inversion in the channel and
allow a current to flow. At low temperature, the sub-threshold
conductance shows series of conductance peaks due to electron
tunneling through a potential well in the conduction band
\cite{sellier-06-cm}. This well located in the channel is
separated from the highly-doped contacts by energy barriers in the
access regions due to a poor n-type doping caused by masking
silicon nitride spacers next to the gate (Fig.\,\ref{fig1}b). Here
we focus on short devices with 60\,nm gate length where the
subthreshold regime exhibits specific conductance peaks due to
isolated dopants in the channel. All the measurements presented
here were gathered on the same sample, but similar features have
been observed in other samples of the same length (the width,
385\,nm here, is not important because of a corner-effect
\cite{fossum-03-ieee,corner-effect}). The experiments were carried
out in a dilution fridge at a base temperature of 100\,mK. The
differential conductance was measured with a lock-in technique
using a 50\,$\mu$V amplitude ac voltage.

The energy barrier between source and drain, measured by fitting
the thermally activated current at temperatures above 150\,K
\cite{sellier-06-cm}, is shown on Fig.\,\ref{fig1}c. At low gate
voltage, when the middle of the channel is the highest point of
the potential profile, the barrier height decreases linearly with
a coupling $\alpha=dE_b\,/\,e\,dV_g=0.68$. An extrapolation of
this linear part to zero barrier shows that the conduction band
edge in the channel reaches the Fermi level at 400\,mV gate
voltage, value above which the low temperature conductance is
expected to rise from zero. \textit{However,} several resonances
(a1, b1, a2, b2) are visible at energies below the conduction band
edge (inset of Fig.\,\ref{fig1}c). They come from isolated dopants
in the channel and will be analyzed in details throughout the
paper. Fig.\,\ref{fig1}c also show that residual barriers remain
at larger gate voltage. Located in the access regions, they create
a potential well in the conduction band of the channel, with
Coulomb blockade at low temperature as revealed by the conductance
peaks c1 and c2.

% ------------------------------------------------------------------------
\begin{figure}[b]
\includegraphics[width=8cm,clip,trim=0 0 0 0]{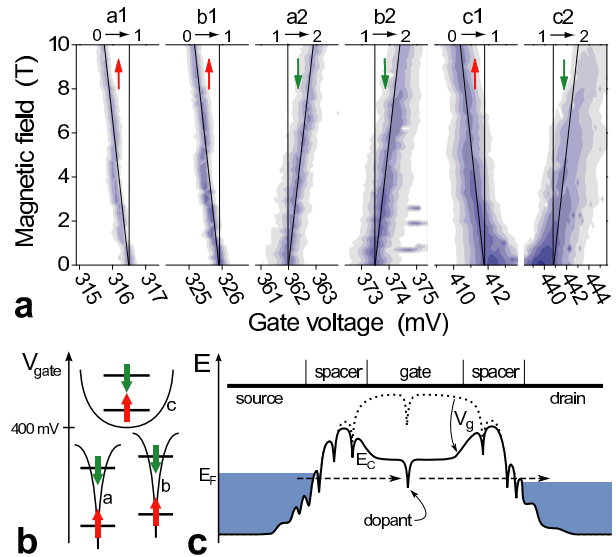}
\caption{(color online). (a) Differential conductance plotted in
color scale (optimized for each peak) showing the peak shifts
under magnetic field and the $g=2$, $S=1/2$ Zeeman shifts (solid
lines) for spin up or down (charging from \emph{0 to 1} or \emph{1
to 2} electrons). (b) Spin polarization of successive resonances
and interpretation in term of two dopants a and b, and one larger
dot c. (c) Conduction band profile with attracting Coulomb
potential of isolated dopants, for an applied gate voltage
corresponding to resonant tunneling through a single dopant in the
channel.} \label{fig2}
\end{figure}
% ------------------------------------------------------------------------

When magnetic field is applied (parallel to the current here),
each peak shifts linearly with field (Fig.\,\ref{fig2}a) and the
slopes are consistent with the Zeeman shift of spins $\pm 1/2$
having a g-factor of 2 as expected for electrons in silicon (using
the conversion factor $\alpha$ of each peak as discussed later).
Peaks labelled with a 1 (2) correspond to tunneling of spin up
(down) electrons under magnetic field. Since the first visible
resonances are always of spin up type (also observed in other
samples), we conclude that the first peaks (a1 and b1) correspond
to tunneling of a first electron ($N=0\rightarrow1$) on two
different impurities (a and b), and the second set of peaks (a2
and b2) correspond to the second charge state ($N=1\rightarrow2$)
of each impurity (see Fig.\,\ref{fig2}b). Note that b1 can not be
the second electron of a1 because a two-electron ground state must
be a singlet with opposite spins \cite{engel-01-prl}. Peaks c1 and
c2 above the band edge are also respectively spin up and down
under magnetic field. They correspond to the first and second
electrons of a larger dot (called c) defined electrostatically by
the gate in the conduction band profile (Fig.\,\ref{fig2}c). In
this sample, the higher charge states of dot c ($N>2$) are not
visible probably due to the progressive loss of Coulomb blockade
above 460\,mV where the confining barriers disappear. In contrast,
in all samples, the impurities like a and b always have exactly
two charge states ($N\leq2$) as expected for dopants because their
single positive charge can bind no more than two electrons and
with a very small second binding energy \cite{hill-77-prl}. In
addition, the measured Zeeman shifts are experimentally
insensitive to the field direction as expected for the 3D Coulomb
potential of dopants as opposed to quantum dots in two-dimensional
electron gases (2DEG).

Isolated n-type dopants in the p-type channel may come from
residual impurities in the SOI layer or from ion implantation of
the contacts. Those contributing to the conductance should be
close to the gate because of the strong band-bending and along the
edges of the wire because this band-bending is stronger due to the
corner-effect \cite{corner-effect}. This strong restriction leads
to a silicon volume containing less than one residual donor
(considering that p-type silicon should contain less donors than
acceptors). The observed isolated donors must therefore arise from
the implantation process.

% ------------------------------------------------------------------------
\begin{figure}[b]
\includegraphics[width=8cm,clip,trim=0 0 0 0]{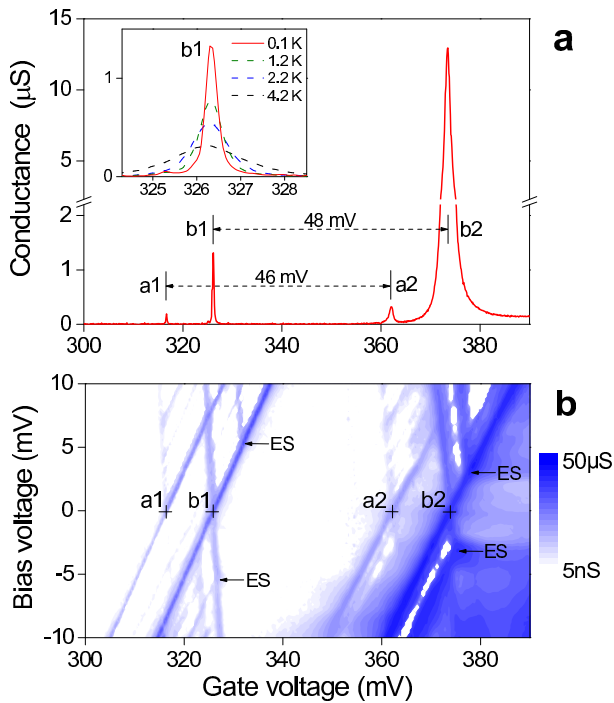}
\caption{(color online). (a) Sub-threshold conductance showing
resonant tunneling through dopants a and b. They have a similar
peak spacing between first and second electron states. Inset: peak
b1 increasing for decreasing temperatures, showing the quantum
regime of Coulomb blockade. (b) Stability diagram, i.e.
differential conductance plotted in color scale versus gate and
bias voltages. Excited states (ES) of dopant b are indicated.
Other weak parallel lines inside the conducting regions result
from local density of state fluctuations in the contacts.}
\label{fig3}
\end{figure}
% ------------------------------------------------------------------------

These isolated dopants are atomic-like systems with one or two
electrons bound by Coulomb attraction to a central ion and can be
characterized by their first and second binding energies. These
values are known for dopants in bulk silicon, but we show in the
following that we can also measure them in gated nanostructures.
Binding energies are related to the gate voltage spacing between
the conductance peaks and the conduction band edge located at
400\,mV (Fig.\,\ref{fig1}c). However the energy change on the
dopant site is a fraction $\alpha=C_g/(C_g+C_s+C_d)$ of the gate
voltage spacing, with $C_s$, $C_d$, and $C_g$ being the
capacitances between the dopant and respectively the source,
drain, and gate electrodes. This ratio is determined from the
conductance diagram versus gate and bias voltages
(Fig.\,\ref{fig3}b) where positive and negative slopes of
conducting sectors equal $C_g/(C_g+C_d)$ and $C_g/C_s$ (the drain
is at ground). We get $C_s=0.15\,C_g$ and $C_d=0.15\,C_g$ for the
first peaks (a1 or b1), and $C_s=0.27\,C_g$ and $C_d=0.31\,C_g$
for the second peaks (a2 or b2) probably due to thinner barriers
at higher gate voltage. The dopants are therefore well centered in
the channel ($C_s \sim C_d$) and mainly coupled to the gate ($C_s
\ll C_g$). The resulting values $\alpha=0.77$ (a1 or b1) and
$\alpha=0.63$ (a2 or b2) are consistent with the coupling factor
$\alpha=0.68$ determined previously from the gate voltage
dependence of the barrier height in the channel below 300\,mV.
This result validates our assumption that the conduction band in
the middle of the channel evolves with the same coupling below and
above 300\,mV. This point is important since we use it to conclude
that the band edge crosses the Fermi level at 400\,mV
(Fig.\,\ref{fig1}c) with two dopants below (peaks a1 to b2) and a
larger dot above (peaks c1 and c2). In the following we use
$\alpha=0.7$ as an averaged value for the conversion factor.

From the 84\,mV (74\,mV) spacing between peak a1 (b1) and the band
edge, we get 59\,meV (52\,meV) for the first binding energy of
dopant a (b). Since this $D^0$ charge state is neutral, its
binding energy is not expected to be modified by the capacitive
coupling with the nearby electrodes. For technological reasons,
both arsenic and phosphorus dopants are used to make the contacts.
Our measured values correspond better to the 54\,meV bulk value of
arsenic than to the 46\,meV of phosphorus \cite{ramdas-81-rpp}.
The small difference between a and b could be related to
fluctuations in the band edge energy in the channel, induced for
example by disorder at the oxide interface.

The binding energy of the second electron (2\,meV only for bulk
dopants \cite{taniguchi-76-ssc}) is usually very small in this
atomic-like system because of a strong repulsive Coulomb
interaction called charging energy (52\,meV for bulk arsenic
dopants). In our case however, since the $D^-$ state is charged,
the capacitive coupling with the nearby electrodes is expected to
reduce the charging energy to a value $e^2/C$ where
$C=C_g+C_s+C_d$. The 46\,mV (48\,mV) spacing between peaks a1 and
a2 (b1 and b2) indicated on Fig.\,\ref{fig3}a corresponds indeed
to a charging energy between the one-electron and the
two-electrons bound states of 32\,meV (34\,meV) lower than the
52\,meV of the bulk. The deduced dopant/gate capacitance
$C_g=3.5\times10^{-18}$\,F ($3.3\times10^{-18}$\,F) corresponds to
a sphere with 2.7\,nm radius in bulk silicon or near a metallic
plane covered with a thin layer of silicon dioxide. This radius
compares very well with the 2.5\,nm Bohr radius of a neutral $D^0$
donor in bulk silicon or with the 3\,nm Bohr radius estimated for
the negatively charged $D^-$ state \cite{norton-76-prl}. As a
conclusion, the second binding energy of the {dopant+gate} system
is significantly increased by the capacitive coupling, and this
effect could provide a longer life-time for the two-electron state
involved in the read-out scheme of dopant-based quantum gates.

% ------------------------------------------------------------------------
\begin{figure}[t]
\includegraphics[width=8cm,clip,trim=0 0 0 0]{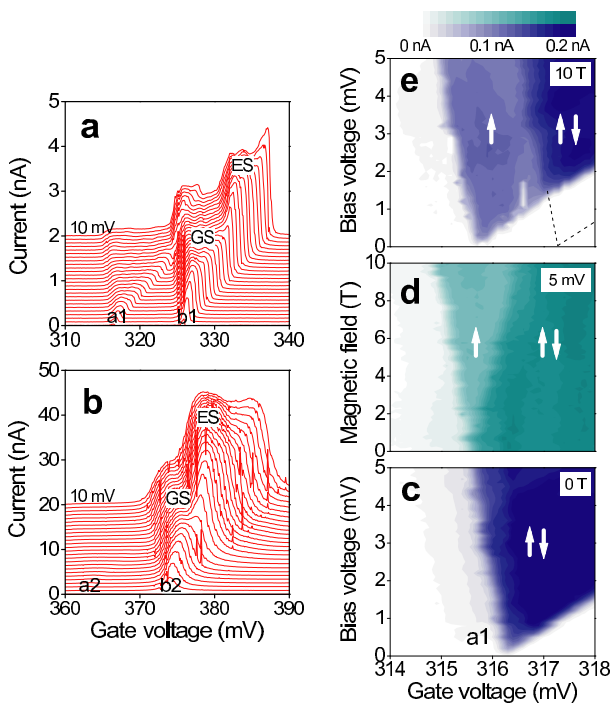}
\caption{(color online). (a)--(b) Current steps versus gate
voltage from 0 to 10\,mV bias. The ground state (GS) and an
excited state (ES) are visible for b1 and b2. (c)--(e) Color-plot
of the current step a1 at different magnetic fields showing the
Zeeman splitting of spin up and spin down contributions. (c)
single step at 0\,T, (d) splitting at 5\,mV bias between 0 and
10\,T, (e) double step at 10\,T with a fully polarized spin up
current step.} \label{fig4}
\end{figure}
% ------------------------------------------------------------------------

We now turn to the excited state (ES) spectroscopy of the two
charge states. In bulk silicon the six-fold valley degeneracy of
the conduction band is lifted on a dopant site by the tetragonal
symmetry and the first electron ground state (GS) is split into
one non-degenerate GS and a few ES all around 12\,meV above the GS
for phosphorus atoms (22\,meV for arsenic) \cite{ramdas-81-rpp}.
An ES of b1 is visible at 5\,meV on Fig.\,\ref{fig4}a as a second
current step on top of the GS one. This lower level spacing than
expected may originate from the strong band bending and Stark
effect near the gate \cite{friesen-05-prl} which are inherent to
these nano-structures. This interesting point will be the object
of further investigations. Regarding the two-electrons state, no
ES would be observed for a bulk $D^-$ ion since any excitation
lies in the continuum \cite{hill-77-prl}. Here however this
charged state is more strongly bound due to the capacitive
coupling with the nearby electrodes and several ES can exist. An
ES of b2 is indeed clearly visible at 3\,meV (Fig.\,\ref{fig4}b)
and has a larger amplitude than the singlet GS in agreement with
the triplet degeneracy expected for this first ES with two
electrons.

These level spacings of several meV are quite large as compared to
2DEG quantum dots and explain why the quantum regime of Coulomb
blockade \cite{beenakker-91-prb} is observed up to 15\,K with
conductance peaks increasing for lowering temperature (inset of
Fig.\,\ref{fig3}a, note that at very low temperature the peak
width is limited by the escape rate to source and drain). Another
consequence of the large level spacing is the absence of level
crossing under magnetic field on Fig.\,\ref{fig2}a because the
1.2\,meV Zeeman splitting at 10\,T is still smaller than the level
spacing. This point is also illustrated on
Fig.\,\ref{fig4}\,c,\,d,\,e showing the current through the state
a1 for a field sweep from 0 to 10\,T. The single current step at
zero field splits into two steps, the left one corresponding to a
flow of spin up electrons and the right one to the sum of spin up
and down. The Zeeman splitting at 10\,T is given by the bias
voltage at which the two spin states contribute and we get 1.2\,mV
as expected for a spin 1/2 electron in silicon, but here the
electron is also bound on a dopant. The dotted lines on
Fig.\,\ref{fig4}e mark the region of the Coulomb-blocked spin down
current. The current in the left region is fully polarized with
spin up electrons and the dopant acts there as a spin filter. A
similar magnetic field behavior has been observed in GaAs/AlGaAs
quantum dots \cite{hanson-03-prl} but with a much lower g-factor
than in silicon and with a possible contribution of orbital
effects absent in case of dopants.

% ------------------------------------------------------------------------

In conclusion, we used silicon nanowires to investigate the
spectrum of single dopants in gated nanostructures. These dopants
form attracting Coulomb potentials and induce resonances in the
sub-threshold conductance due to electron tunneling through their
discrete energy levels. We observed the $D^0$ and $D^-$ charge
states, measured their binding energy and the capacitive coupling
with nearby electrodes in case of charged states. By means of
transport spectroscopy versus bias voltage and magnetic field, we
measured the quantum level spacing and Zeeman spin splitting.

% ------------------------------------------------------------------------

We thank F.~Koppens and M.~Sanquer for useful discussions.

% ------------------------------------------------------------------------

% ------------------------------------------------------------------------

\begin{thebibliography}{30}
\bibitem{corner-effect}
Transport is confined to the edges of the silicon wire because the
conduction band edge is lower due to a stronger electric field
produced by the gate.
\end{thebibliography}
\end{document}